\documentstyle[aps,preprint]{revtex}   
\begin{document}   
\tighten   
\def\a{a}   
\def\n{n}   
\def\H{H}   
\def\D{D}   
\def\m{m}   
\def\t{t}   
\draft   
\preprint{hep-th/0106241}   
\date{June 26, 2001}

\title{\Large\bf  BPS Branes in Cosmology}   
\author{Renata Kallosh,$^1$  Lev Kofman,$^2$   Andrei Linde,$^1$   
and Arkady Tseytlin$^{3}$ \footnote{ Also at  Imperial College, London and   
Lebedev Physics Institute, Moscow. }} \address{$^1$ Department of Physics,   
Stanford University, Stanford, CA 94305-4060, USA} \address{$^2$ CITA,   
University of Toronto, 60 St. George Street, Toronto, ON M5S 3H8, Canada}   
\address{$^{3}$   Department of Physics,   
The Ohio State University, Columbus, OH 43210, USA }   
 \maketitle

\begin{abstract}   
The possibility to study the M/string theory cosmology via 5d bulk \& brane action is investigated. The role of the 4-form field in the theory of BPS branes in 5d is clarified. We describe arguments suggesting that the effective 4d description of the universe in the ekpyrotic scenario (hep-th/0103239) should lead  to contraction rather than expansion of   
the universe. To verify these arguments, we study the full 5d action prior to its integration over the 5th dimension. We show that if one adds the potential $V(Y)$ to the action of the bulk brane, then the metric ansatz used in the ekpyrotic scenario does not solve the dilaton and gravitational equations. To find a consistent cosmological solution one must use a more general metric ansatz and a complete 5d description of the brane interaction instead of simply adding an effective 4d bulk brane potential $V(Y)$.   
\end{abstract}   
\pacs{SU-ITP-01/32 \hskip 2.6 cm CITA-01-14   \hskip 2  cm OHSTPY-HEP-T-017}

\section{Introduction}

During the last few years there were many  attempts to construct a   
consistent   brane cosmology, see e.g.   
\cite{brane,Binetruy:2000ut,DvaliTye} and references therein. One of the   
most interesting  possibilities is to use supersymmetric BPS branes in   
cosmology.  Studies of this idea  developed from the no-go theorems   
for nonsingular supersymmetric domain walls \cite{KL} to the construction   
of supergravity in singular spaces \cite{BKV} where the bulk and brane   
actions are supersymmetric. Investigation of the BPS brane cosmology, i.e.   
the theory of interacting and moving near-BPS branes, has brought an   
additional level of complexity, both on the technical and on the   
conceptual level. 

One of the most challenging  recent  attempts  to construct 
a consistent  cosmology  based on  a picture of colliding BPS branes 
is  ekpyrotic scenario  \cite{KOST}.
In this paper we will discuss some of the general issues   
of the BPS brane cosmology  by critically analyzing  the  ekpyrotic   
scenario.   
   
It was claimed  in  \cite{KOST}  that the ekpyrotic scenario is based on the   
Ho\u{r}ava-Witten (HW) phenomenology,  and it solves all major   
cosmological problems without using inflation \cite{KOST}. However, in   
\cite{KKL} it was argued that these claims were overly optimistic. First   
of all, the standard HW phenomenology (including most of its versions with   
non-standard embedding) is based on the assumption that we live on the   
positive tension brane \cite{HoravaWitten}. Meanwhile, the model proposed   
in \cite{KOST} was based on an unconventional approach to HW theory,   
assuming that we live on the negative tension brane. This required a   
substantial reformulation of HW phenomenology   
\cite{Donagi:2001fs,Khoury:2001iy}. In particular, one must revise  the   
standard assumption \cite{HoravaWitten} that the gauge coupling on the   
hidden brane is large.  In \cite{KKL} a different version of this scenario   
was proposed, assuming that we live on the positive tension brane, in   
accordance with \cite{HoravaWitten}. It was called pyrotechnic 
scenario.\footnote{In the pyrotechnic scenario, unlike in the 
ekpyrotic scenario, we do not make any attempts to avoid inflation. 
In fact we have argued in \cite{KOST} that avoiding inflation in this 
scenario may require additional fine-tuning.}   
We will discuss this in Section \ref{HW} of our paper and explain   
that the relevant issue is not the standard versus non-standard embedding,   
but Ho\u{r}ava-Witten phenomenology \cite{HoravaWitten} versus   
Benakli-Lalak-Pokorski-Thomas \cite{Benakli:1999sy} phenomenology.   
   
Other concerns include extreme fine-tuning of the brane potential $V(Y)$   
required in the ekpyrotic scenario. In particular, the potential must be   
extremely small (suppressed by a factor at least as small as $10^{-50}$)   
near the hidden brane. This makes it very difficult to understand how this   
scenario could be made consistent with the brane stabilization \cite{KKL}.   
To solve the homogeneity problem in this scenario, one would need the   
branes from the very beginning to be parallel to each other with an   
accuracy better than $10^{-60}$ on a scale $10^{30}$ times greater than   
the distance between the branes. In our opinion,  these problems, as well   
as several  other problems of the ekpyrotic scenario pointed out  in   
\cite{KKL}, remain unresolved.   
   
In this paper we would like to consider some other aspects of the   
ekpyrotic scenario. First of all, it was claimed in \cite{KOST} that the   
action   and the cosmological solution describing three static branes in   
the ekpyrotic scenario have been obtained in \cite{lukas,universe}.   
However, the 5d bulk \& brane action in \cite{universe} was found for the   
case of  the two boundary branes only. We will show that it cannot be   
generalized to the case of boundary and bulk branes using the formalism of   
Ref. \cite{lukas,universe}.   
   
To add the bulk brane to this construction  one must use the 4-form field,   
which was introduced in the context of 5d supergravity in singular spaces   
by Bergshoeff, Kallosh and Van Proeyen \cite{BKV} and recently generalized   
in \cite{Fujita:2001bd}. We will show that the corresponding part of the   
5d action and the part of the solution given in \cite{KOST} are not quite   
correct as far as coefficients are concerned; we will present the   
corrected  action and  static BPS solution describing boundary and bulk   
brane in Section \ref{MS} (some technical details are given also in Appendix).

In Section \ref{4d} we will discuss an effective 4d description of the 5d   
cosmology and give an argument that the 4d space in the ekpyrotic scenario   
can only collapse.\footnote{Various versions of this argument were   
suggested to us independently by  T. Banks, G. Dvali, and  J. Maldacena   
\cite{all}.} According to \cite{KOST}, this is indeed the case. The scale   
factor of the universe, which was obtained after integrating the 5d action   
over $y$ (the 5th dimension) and solving equations in the effective  4d   
theory, decreases. The authors of \cite{KOST} argued that one can go back   
from the effective 4d theory to 5d theory and show that the scale factor   
of the visible brane grows. We believe, however, that in order to go back   
from the effective 4d description to the 5d theory  one should perform an   
explicit investigation of the 5d geometry before the integrating over the   
5th dimension.   
   
This is not an easy task since the bulk brane potential $V(Y)$ driving   
expansion of the universe in the ekpyrotic scenario was added  in   
\cite{KOST} by hand to the 4d formulation of the theory, rather than to   
the 5d theory. One can only speculate how one should interpret this term   
from the point of view of the 5d theory.   
   
In \cite{KOST}  this term was interpreted as a correction to the bulk   
brane action, which means that it represents an effective delta-functional   
addition to the total energy concentrated on the bulk brane. We will show   
in Sect. \ref{5d} that in this case  the time-dependent  ansatz for the metric and fields used in  \cite{KOST} does not provide a consistent solution to the 5d gravitational equations  and to the equation for the dilaton field. In such a situation it is dangerous to take averages of the 5d action or of the 5d equations over $y$. If one does so with the 5d dilaton equation, one finds that the whole universe, including the visible brane, should exponentially collapse rather than expand.   
   
In order to find a consistent solution of 5d equations one needs to make   
at least two important modifications discussed in Section \ref{VY}. First   
of all, when one considers moving branes, or branes having energy momentum   
tensor different from the 4d cosmological constant, one should use a   
general ansatz for the metric and for the fields   which respects the   
(planar) symmetry in the problem \cite{Binetruy:2000ut} rather than the   
factorized  ansatz  used in \cite{KOST}. The typical situation there when   
the junction conditions on the branes are taken into account is that the   
5d metric has a non-factorizable dependence on time and fifth direction.   
   
Moreover, we believe that the brane potential $V(Y)$ should be interpreted   
as the ``radion potential.'' It represents  the total energy of the   
configuration involving the bulk brane standing at the distance $Y$ from   
the visible brane \cite{GW}. In such a situation it may be incorrect to   
represent $V(Y)$ as a delta-functional contribution to the energy density   
localized on the bulk brane. Instead of that, one should   find out the   
distribution of the fields responsible for the emergence of the long range 
interaction described by the potential $V(Y)$, and   
substitute their $y$-dependent energy-momentum tensor (rather than $V(Y)$)   
into the 5d equations. A simplest example of a similar situation is given   
by the energy of an electric  capacitor. The energy of interaction of the   
charged plates of a capacitor $V(R)$ is proportional to the distance $R$   
between the plates. However the electrostatic energy density $\sim E^2$   
is concentrated not at the plates but in the bulk.   
   
We conclude that in addition to many other problems of the ekpyrotic   
scenario discussed in \cite{KOST}, there exists another one. The ansatz   
for the metric and fields used in \cite{KOST} does not provide a correct   
solution to the dilaton equation and the 5d Einstein equations.

\section{\label{HW}Brane tension and HW phenomenology}

According to the Ho\u{r}ava-Witten theory \cite{HoravaWitten,universe},   
the universe consists of two branes in  5d space, which appeared after 6   
dimensions of the  11d space were compactified on Calabi-Yau (CY) space.   
The unification of weak, strong and electromagnetic interactions is   
achieved on the visible  brane, with the gauge coupling  $\alpha_{GUT}   
\sim 0.04$. The standard Ho\u{r}ava-Witten phenomenology   
\cite{HoravaWitten}, including its versions with non-standard embedding   
\cite{Lukas:1999hk}, is based on the assumption that we live on   
 the positive tension   
brane, and the volume of the CY space decreases towards the hidden brane,   
whereas the gauge coupling constant increases. This leads to the strong   
gauge coupling on the hidden brane, $\alpha_{\rm hid} = O(1)$. In the   
strong coupling regime one can obtain gaugino condensation, which plays an   
important role in the HW phenomenology.   
   
The ratio of the gauge couplings is inversely proportional  to the ratio   
of Calabi-Yau volumes at the positions of the branes \cite{HoravaWitten},   
\begin{equation}   
  { \alpha_{\rm hid}} = \alpha_{\rm GUT} {V_{\rm vis} \over V_{\rm   
  hid}} \ .   
\label{couplings}   
\end{equation}   
The visible brane is located at $y=0$ and the hidden one at $y=R$ so that   
$ V_{\rm vis}= e^{\phi (0)}$ and $V_{\rm hid}= e^{\phi(R)}$. This formula   
is valid for the versions of HW theory with the standard embedding without   
M5-branes, as well as for the versions with the non-standard embedding   
with M5-branes present in the bulk, see e.g. a discussion of the   
phenomenology of the theories with the non-standard embedding by Lukas,   
Ovrut and Waldram \cite{Lukas:1999hk}.

In  \cite{KOST,Donagi:2001fs} it was assumed that the  tension of the   
visible brane $-\alpha$ is negative. The volume of the CY space in   
\cite{KOST} is proportional to $D^3(y)$, where $D(y) = C + \alpha y$. The   
function $D(y)$ grows at large $y$. As a result, the value of $D^3(y)$   
near the hidden brane is 50 times greater than near the visible brane for   
the parameters used in \cite{KOST}; it is 27 times greater for the   
parameters chosen in the replaced version of their paper. The authors of   
the ekpyrotic scenario  did not calculate the value of $\alpha_{\rm hid}$   
in their scenario \cite{KOST,Donagi:2001fs}, so we will estimate it now.   
If one has $\alpha_{GUT}  \sim  0.04$ on the visible brane, one finds   
$\alpha_{\rm hid} = O(10^{-3})$. This is way too small to lead to the   
usual HW phenomenology, which requires $ \alpha_{\rm hid} \sim 1$ and   
gaugino condensation \cite{HoravaWitten}. We are not saying that a   
consistent phenomenology  with $\alpha_{\rm hid} \ll \alpha_{GUT}$ is   
impossible, see e.g.  \cite{Benakli:1999sy},  but this is a rather   
unconventional possibility.

Thus, we believe that the original version of the ekpyrotic scenario was   
at odds with the standard HW phenomenology as defined in   
\cite{HoravaWitten}. So why was the tension of the visible brane chosen to   
be negative in \cite{KOST}, despite all complications associated with such   
a choice? In the original version of Ref. \cite{KOST}  we read: {\it As we   
will see in Section VB,   it will be   necessary  for the visible brane to   
be in the small-volume region of space-time.}  This statement, as well as   
the related conclusion that the spectrum of perturbations in the ekpyrotic   
scenario  is blue, was  emphasized in many places of the text. The reason   
of this statement, as explained in Section VB of \cite{KOST}, was rooted   
in the idea that the decrease of $D(Y)$ is required for generation of   
density perturbations in the ekpyrotic scenario.   
   
In ref. \cite{KKL} a  simple description of generation of density   
perturbations in the ekpyrotic scenario was presented\footnote{   
Recently it was claimed \cite{Lyth:2001pf} that if one takes into   
account gravitational backreaction, no perturbations of metric are   
produced in the ekpyrotic scenario. If this is the case, there is no need   
to continue discussion of this scenario.  However, it is not obvious to us   
that the absence of fluctuations of the effective 4d metric discussed in   
\cite{Lyth:2001pf} implies the absence of  the 5d metric   
perturbations and the absence of  the time delay of the brane collision.   
Perturbations investigated in \cite{KOST,KKL} appeared not because of  the   
fluctuations of the curvature of the 4d spacetime prior to the brane   
collision, but because of the ``radion" perturbations $\delta Y$ (related   
to  $\delta g_{yy}$). These perturbations, 
describing an inhomogeneous embedding  of the bulk brane in the 5d spacetime,   
cannot be reduced to perturbations of the 4d geometry.   
In any case, before studying perturbations, one should first carefully   
examine the behavior of the nonperturbed metric. This is what we are going   
to do in our paper.} using the methods developed in the theory of   
tachyonic preheating \cite{tach}. It was  shown in \cite{KKL}, in   
particular, that the requirement that the visible brane must be in the   
small-volume region of space-time is not necessary. Thus there is no   
reason to abandon the standard HW phenomenology and assume that we live on   
the negative tension brane. An improved version of the ekpyrotic scenario   
based on the assumption that we live on the positive tension brane was   
called   ``pyrotechnic universe'' \cite{KKL}.

\section{\label{MS} M/string theory and 5d BPS domain walls}   
   
Compactification of M/string theory with extended objects down to 5d   
sometimes leads to appearance of supersymmetric domain walls. It has been   
explained in \cite{BKV} that the supersymmetric domain walls in 5d must be   
charged. The relevant  4-form $A$ and 5-form field strength $F=dA$ play an   
important role in the supersymmetry  of the bulk \& brane construction of   
\cite{BKV}. It is the consequence of the M/string theory supersymmetry   
where RR-fields and M-theory form-field provide the balance of forces   
between BPS extended objects.

Ekpyrotic scenario is based on 11d  theory with Ho\u{r}ava-Witten domain   
walls and some M5-branes between them. A compactification of  this system   
  may lead to  5d theory with   
charged 3-branes. A complete M-theory  derivation of the  5d theory with   
boundary HW domain walls and bulk branes between them was not actually   
worked out. Only the  part of it with 2 HW walls was reduced to 5d in   
\cite{universe}. The 4-form field describing  HW domain walls and   
compactified M5-branes was introduced recently in \cite{KOST} and it was   
argued in \cite{Khoury:2001iy} that the 4-form formulation of the action   
in \cite{KOST} is equivalent to the action presented in \cite{universe}.   
It was claimed that  this  is easily seen by eliminating the 4-form using   
its equation of motion.   
   
Since the purpose of these notes is to find the correct equations in 5d   
which are related to M/string theory and BPS construction, we have   
examined this claim and found that it is not quite correct. We use in this   
analysis  the supersymmetric 5d  theory  with the 4-form field introduced   
in \cite{BKV}.   
\begin{itemize}   
   
\item  In \cite{universe} there are only 2 HW walls, no branes in between.   
We will show that in case when   
 in addition to the  2 boundary branes  there are bulk branes   
 present, the   
covariant action with the bulk dilaton potential (replacing  the   5-form   
contribution) of the type \cite{universe} does not exist. Only the   
formulation with the 4-form field  can describe the two boundary branes   
and a brane in between. We refer the reader to the Appendix A where all   
technical details are explained.   
   
\item In case of only 2 HW walls the action in \cite{KOST} is not   
equivalent to the action in \cite{universe} unless the factor in front of   
the ${\cal F}^2$ kinetic term is changed.   
   
\item   
 The expression for the  4-form  for the static solution   
in \cite{KOST}   
   is not correct. The numerical factor has to be   
 modified and the sign has to be changed to make it a BPS solution.   
   
\end{itemize}   
We start with the corrected  version of the action in \cite{KOST} (note   
the factor $3/2$ in front of ${\cal F}^2$)   
\begin{eqnarray}   
 & & S=\frac{M_5^3}{2}\int_{{\cal M}_5}   
d^5x\sqrt{-g}\left(R-\frac{1}{2}(\partial\phi)^2-{3\over   
2}\frac{e^{2\phi}{\cal   
F}^2_5}{5!}\right) \\   
& &-3\sum_{i=1}^3\alpha_iM_5^3\int_{{\cal M}_4^{(i)}}   
d^4\xi_{(i)}\Bigl(\sqrt{-h_{(i)}}e^{-\phi}   
 -\frac{\epsilon^{\mu\nu\kappa\lambda}}{4!}{\cal   
A}_{\gamma\delta\epsilon\zeta}   
\partial_{\mu}X^{\gamma}_{(i)}\partial_{\nu}X^{\delta}_{(i)}   
\partial_{\kappa}X^{\epsilon}_{(i)}\partial_{\lambda}X^{\zeta}_{(i)}\Bigr) \ ,   
\label{eq:5daction}   
\end{eqnarray}   
where $\alpha_1= -\alpha $, $\alpha_2= \alpha-\beta$, $\alpha_3=\beta$.   
The corrected form of the static solution is\footnote{Here $y$ is a point   
of $S^1/Z_2$, i.e. $-R < y \leq  R$, with  $0$ and $R$  as fixed points,   
$R$ identified with $-R$. This explains the factor of 2 for the   
fixed-point brane sources   at $0, R$  (accounting for their images), and   
the presence of  the brane in the bulk at $y=Y$ and its image at $y=-Y$.}   
\begin{eqnarray}   
& & ds^2=D(y)(-N^2dt^2+A^2d\vec{x}^2)+B^2 D^4(y)dy^2 \ ,   
\label{eq:static1}   
\\ & & e^{\phi}=B D^3(y) \ , \label{eq:four}   
 \\ & & D(y)=\alpha y +C   
\;\;\;\;\;\;\;\;\;\;\;\;\;\;\;\;\;\;\;\;\;\;\;\;\;\;{\rm   
for}\;\;y<Y, \\   
& & \;\;\;\;\;\;\;\;\;=(\alpha-\beta)y+C+\beta Y \;\;\;\;\;\;\;{\rm   
for}\;\;y>Y, \label{eq:static}   
\end{eqnarray}   
where $A,B,C,N$  are constants, $C>0$,   
\begin{equation}   
{\cal A}_{0123}= + A^3 N B^{-1} D^{-1}(y)      \ ,  \ \ \ \ \ \ \ \   
             {\cal F}_{0123y}= - A^3  N B^{-1} D^{-2}(y) D'(y) \ ,   
 \label{correct}   
\end{equation}   
and   
\begin{equation}   
 [D(y)]^{''}= 2 [\alpha\delta (y)-(\alpha -\beta )   
  \delta (y-R)] -\beta \delta (y-Y) -\beta \delta (y+Y) \ .   
\label{D}   
\end{equation}   
 The factor $-A^3 N   
B^{-1}$  in   $\cal F$ in   eq. (\ref{correct})  was   
 absent in \cite{KOST}.   
 The sign  of ${\cal A}_{0123}$ and ${\cal F}_{0123y}$  (which differs from  the one in \cite{KOST}),   
is easily checked  by observing  that the  force on a static probe brane   
parallel to the source branes must vanish.\footnote{ The  world-volume   
term   
 cancels against   the  Wess-Zumino term in the static probe brane  action. We are assuming the standard   
convention $\epsilon^{0123}=+1$. }   
   
The static BPS solution is valid for branes that are not moving. It  may   
serve as a starting point for finding time-dependent cosmological   
solutions.

\section{ 4d view on ekpyrotic universe}\label{4d}   
Before investigating time-dependent cosmological solutions in 5d, let us   
see what could be expected from the point of view of the effective 4d   
theory. Indeed, if one considers the situation when the distance between   
the branes is very small and their motion is slow, one could expect that   
in the first approximation it should be possible to describe low energy   
theory from the point of view of  4d Einstein gravity (or Brans-Dicke   
theory) coupled to matter. Deviations from this description could occur if   
there were some processes with the energies comparable to the inverse   
distance between the branes. However, in the ekpyrotic scenario all energy   
scales  (reheating temperature, Hubble constant, etc.) are several orders   
of magnitude smaller than the inverse distance between the branes $1/R$.   
   
Therefore  one may expect that the ekpyrotic scenario can be described   
entirely in terms of the 4d theory. But in such a case the universe, which   
was static in the beginning of the process, can only collapse. We present   
here, in a slightly modified form,  the basic argument of Ref. \cite{all}; 
see also \cite{KKL}.

Let us write down the Einstein equations for a homogeneous flat universe.   
The first equation is   
\begin{equation}\label{fried1}   
H^2 = {8\pi G\over 3} \rho \ .   
\end{equation}   
The second equation, which follows from the first one and the energy   
conservation, can be represented in the following form:   
\begin{equation}\label{fried2}   
\dot H  = - 4\pi G (\rho + p) \ .   
\end{equation}   
Here $\rho$ and $p$ are  the  density and pressure in the effective 4d   
theory, and  $H = {\dot a/a}$,  where  $a$ is the scale factor in the 4d   
space on the visible brane.   
   
In the beginning, the branes do not expand, $H= 0$, and the effective   
energy density and pressure vanish for the static brane configuration   
considered in the previous section.   
   
This situation changes when one adds by hand the potential energy $V(Y)$   
associated with the position of the bulk brane. According to \cite{KOST},   
$V(Y)< 0$. This, however, would be inconsistent with Eq. (\ref{fried1})   
unless one  assumes that the bulk brane has nonzero velocity from the very   
beginning, so that the total energy density is non-negative. We do not   
want to speculate on how this configuration could emerge; see Ref. \cite{KKL}   
for a discussion of the problem of initial conditions in the ekpyrotic   
scenario.   
   
What is more important,   Eq. (\ref{fried1}) implies that $\dot H \leq 0$   
because $\rho + p \geq 0$ in accordance with the null energy   
condition. Thus, if the universe begins in a static state, $H = 0$, then   
it can only collapse, since $H = \dot a/a \leq 0$. Therefore one may   
argue that the ekpyrotic scenario cannot describe an expanding universe   
\cite{all}.   
   
Let us compare these expectations with the results obtained in   
\cite{KOST}. To describe the motion of the bulk brane in the ekpyrotic   
scenario the authors started with  the  factorized ansatz based on   
~(\ref{eq:static1})--(\ref{eq:static}) where it was assumed  that some of the   
parameters of the static solution become functions of time but not of   
$y$\, ($A,N, Y \to A(t), N(t), Y(t)$), whereas some other parameters   
remain constant ($\dot B =\dot C=0$). They  substituted  this modified   
ansatz into the action~(\ref{eq:5daction}), and {\it integrated over $y$}.   
In this way they obtained the 4d ``moduli space'' action  with the   
Lagrangian density ${\cal L}= {\cal L}_{bulk}^{4d}+{\cal L}_{\beta}\ , $   
where   
\begin{equation}\label{bulklagr}   
 {\cal L}_{bulk}^{4d} = -2 \frac{3 A^3 B M_5^3}{N} \int _{0}^{R}dy D^3 (y, Y) \left[   
 \left(\frac{\dot{A}}{A}\right)^2 + 3 \frac{\dot{A}}{A}\frac{\dot   
D}{D} +{1\over 2}\left (\frac{\dot D }{D}\right)^2 \right] ,   
\end{equation}   
and   
\begin{equation}   
  {\cal L}_{\beta} = \frac{3\beta M_5^3   
A^3B}{N}\, \left[\frac{1}{2}\D^2(Y)\dot{Y}^2\right]  . \label{branelagr}   
\end{equation}   
We gave here a form of the  Lagrangian in which it is clear that it is   
 integrated over the full range of the 5th coordinate.

In \cite{KOST} the notation $I_k(Y)\equiv \int _{-R}^{R}dy D^k(y, Y)=   
2\int _{0}^{R}dy D^k(y, Y)$ ($k=1,2,...$) and the explicit form of $D$ was   
used so that $\dot D= 0$ at $|y| < Y$ and $\dot D= \beta \dot Y$ at $|y|   
\geq Y$. After obtaining the effective 4d action by integration over $y$,   
the authors added by hand an important term, which plays a crucial role in   
their scenario:   
\begin{equation}\label{add}   
\Delta{\cal L}_{\beta}=- 3\beta M_5^3 A^3B N\, V(Y) \ .   
\label{eq:4daction}   
\end{equation}   
The effective potential $V(Y)$  may appear, e.g., as a result of the   
nonperturbative effects associated with open M2-brane instantons   
\cite{Moore:2000fs}. It was assumed in \cite{KOST} that $V(Y) \sim e^{-   
\alpha m Y}$ at large $Y$, and that it vanishes at $Y = 0$.   
 Here   
 $m$ is  some positive numerical constant specified  (together with the tension $\alpha$)   
in \cite{KOST}.

The next step was to replace $A$ and $N$ by $a = A(BI_3M_5)^{1/2}$ and $n   
= N(BI_3M_5)^{1/2}$. This  gives the effective 4d theory in the following   
form:   
\begin{eqnarray}   
 {\cal L} =\frac{3\a^3M_5^2}{\n}   
\left \{-\left(\frac{\dot{\a}}{\a}\right)^2 +J(Y) \beta^2\dot{Y}^2   
+\frac{\beta}{I_3}\left[\frac{1}{2}D(Y)^2\dot{Y}^2   
-\n^2\frac{V(Y)}{BI_3M_5}\right]\right \}, \label{4daction}   
\end{eqnarray}   
where   
$J(Y)\equiv\left(\frac{9I_{2b}^2}{4I_3^2}-\frac{I_{1b}}{2I_3}\right)$ and   
$D(Y)= \alpha  Y + C $. Then the authors of \cite{KOST} neglected the   
terms proportional to $\beta^2$ (we will return to this point in the next   
section) and studied the 4d theory with the effective Lagrangian   
\begin{eqnarray}   
 {\cal L}_{\rm eff} =\frac{3\a^3M_5^2}{\n}   
\left \{-\left(\frac{\dot{\a}}{\a}\right)^2   
 +\frac{\beta}{I_3}\left[\frac{1}{2}D(Y)^2\dot{Y}^2   
-\n^2\frac{V(Y)}{BI_3M_5}\right]\right \}. \label{eff}   
\end{eqnarray}   
Since this Lagrangian leads to the equations which look exactly like the   
usual Friedmann equations for the scale factor $a$, the authors obtained   
the result which we expected on the basis of our general arguments: $H =   
{\dot a\over a} < 0$, i.e the 4d universe {\it contracts}.   
   
At this stage it is very important to realize that we are not talking here   
about contraction of the bulk brane, or the visible brane, or the hidden   
brane. We are investigating an effective 4d geometry where  the   
distinction between these branes completely disappeared. It was ``washed   
away'' by  the integration over $y$.  In a certain sense, one may imagine   
that integration over $y$ ``glues'' the three branes together. Thus we are   
talking about the contraction of the whole 4d space rather than of  the   
one of the branes. All the difference between the branes in 4d must be   
encoded in the dynamics of the moduli fields, but not in the different   
rate of expansion or contraction of different branes.   
   
Note that the contraction of the universe was crucial for ekpyrotic   
scenario in order to gain a residual kinetic energy of the moving brane   
and transfer it to the radiation upon the collision. This residual  energy   
was suppressed by the small coefficient $O(\beta/\alpha)$ as compared to   
the maximal kinetic energy of the brane.   
   
So where did the expansion of the universe come from in  \cite{KOST}?   
After solving the effective 4d equations, the authors decided to   
``unglue'' the branes and go back to 5d. They studied the difference   
between the expansion of the universe as seen by different observers   
living on different branes and concluded that whereas the universe   
described by the overall scale factor $a$ collapses, the visible brane,   
described by the scale factor $a_{-}$, may expand.  

It is this point that is in an apparent contradiction with our 4d   
expectations suggesting that the scale factor of the universe in the   
ekpyrotic scenario cannot expand. However, the statement that $a_-$   
expands  was based on the assumption that  the function $A$ in the metric   
(\ref{eq:static1}) depends only on $t$ but not on $y$, and that the   
effective 4d  description with the metric ansatz of \cite{KOST} provides a   
correct solution for the scale factor of the universe in the full 5d   
theory. As we will see in the next section, our analysis of the 5d   
solutions does not confirm this assumption.

\section{5d theory in the bulk}\label{5d}   
   
An assumption  of the ekpyrotic cosmology  in \cite{KOST} was that the   
time dependent solution can be obtained from the static solution using the   
``moduli space'' approximation, i.e. by replacing the constant  moduli   
parameters of the static solution by time-dependent functions. More   
precisely, the ansatz used in \cite{KOST} was to introduce only time  {\it   
but not $y$} dependence into $A,N,Y$ which were constants in the  static   
BPS solution, so that they become  $A(t)$, $N(t)$, $Y(t)$. All dependence   
on time in $D(y)$ then enters only via $Y(t)$, i.e. for the cosmological   
solution one takes $D=D(y, Y(t))$.  The two extra parameters $B,C$  which   
appeared in the metric, in the dilaton and in the 4-form field  were   
assumed to be time-independent.

As already discussed above, in  \cite{KOST}  this ansatz was  plugged   
into the  5-d action,   
 and then integration  over $y$  gave an  action  for time-dependent functions   
only. That action was taken as a starting point for a cosmological   
analysis. However, it is not clear a priori why this procedure is actually   
{\it consistent}, i.e.  why the solutions of the resulting effective   
equations represent at the same time the solutions of the  original {\it   
5-dimensional}  gravitational equations.   
   
Indeed, it is well-known (see e.g. \cite{ModCosm}) that the  moduli space   
approximation, i.e. replacing moduli  by time-dependent function may not   
always be a consistent. Unless  the  dependence on the internal coordinate   
($y$ in the present case)  is  ``homogeneous''  so that  it effectively   
``scales out''   of the  time-dependent higher-dimensional equations,  one   
cannot simply replace these equations by their  integrated (averaged over   
$y$) version --  the moduli space approximation ansatz  will not be   
consistent with the full set of the gravitational equations.

The question now arises whether it is possible to set up a 5d theory of   
ekpyrotic cosmology before integrating over $y$.   
   
Since in \cite{KOST} the interaction between the branes in 5d bulk was not   
specified but only the potential $V(Y)$ which lives on the bulk brane at   
$y=Y$ was introduced, we will also take this ansatz as part of the   
definition of the 5d theory. As we shall see, this will lead  to  a  major   
problem with ekpyrotic cosmology, a contradiction with 5d solutions of   
equations of motion.   
   
It could be expected that the   5d theory (\ref{eq:5daction}) does not   
describe the physics of the brane collision and does not address the   
moduli stabilization problem \cite{KOST}.  The problem we are discussing   
now is completely different. Here we will analyse equations in the bulk   
during  the roll of the bulk brane until it reaches the minimum of   
its potential $V(Y)$. If the 5d equations do not work even at this stage,   
then we do not have a consistent scenario not only during the brane   
collision but even before it.   
   
To make the discussion as clear as possible, we shall start with the   
simplest equation -- the one for the dilaton $\phi$. The dilaton equation   
of motion in the bulk (away from the branes where the source terms vanish)   
is given by   
\begin{equation}   
 {2\over M_5^3} {\delta S_{bulk}\over \delta \phi(t,y)}=   
\partial_\gamma(\sqrt{-g}\, g^{\gamma\delta}\,\partial_\delta \phi)-3\sqrt{-g}\,\frac{e^{2\phi}{\cal   
F}^2}{5!}=0 \ . \label{dilaton}   
\end{equation}   
It is satisfied for the above static solution (\ref{eq:static1})-(\ref{D}).   
Plugging in the time-dependent ansatz of \cite{KOST}, we find that the new   
term in the dilaton equation, which was  absent in static case, is simply   
the time-derivative one ($\dot D \equiv \partial_t D$):   
\begin{equation}   
\partial_t(\sqrt{-g}\, g^{tt}\,\partial_t \phi) =- 3 \partial_t(A^3 D^2 B   
N^{-1} \dot D) \ .   
 \label{test}   
\end{equation}   
The other terms in the dilaton equation, $   
\partial_y(\sqrt{-g}\, g^{yy}\, \partial_y \phi)-3\frac{e^{2\phi}{\cal   
F}^2}{5!}$, cancel not only for the static ansatz but for the time   
dependent ansatz as well, for the corrected action and solution given in   
(\ref{eq:5daction})-(\ref{correct}). Following \cite{KOST} in assuming   
that $B$ and $C$ are constant, we may study the dilaton equation which   
then reduces to the condition that (\ref{test}) should vanish for all   
values of $y$ away from the branes. For  $y< Y$ the term  (\ref{test})   
vanishes since for $D= C+\alpha y$ we find that   $\dot D=0$. However, for   
$y$ behind the moving brane, i.e. $y> Y$, we find that $\dot D= \beta \dot   
Y \not = 0$, and therefore the  correction to the dilaton equation due to   
the time evolution does not vanish automatically. In this case Eq.   
(\ref{dilaton}) looks as follows:   
\begin{equation}   
 -{2\over 3 M_5^3} {\delta S_{bulk}\over \delta \phi}= \partial_t(A^3 D^2 B N^{-1}   
\beta \dot Y) =0 \ ,  \qquad y> Y(t) \ .\label{testbehind}   
\end{equation}   
 If we assume as  in \cite{KOST} that $B$ is constant and   
$N(t)=A(t)$ (which corresponds to the  choice of  $t$ as a conformal   
time), we get the condition   
\begin{equation}   
 \partial_t (A^2 D^2 \dot Y)=  0, \qquad  y >Y(t) \ . \label{test1}   
\end{equation}   
The solution of this equation is   
\begin{equation}   
A^2(t)\, D^2(y, Y(t))\, \dot Y(t)=  f(y) \ ,  \qquad y >Y(t) \ ,   
\label{constraint}   
\end{equation}   
where  $f(y)$   is an arbitrary function of $y$  that does not depend on   
$t$.   
   
Let us first look at this equation in spirit of \cite{KOST}, replacing   
$D^2(y, Y(t))$ in the first approximation by its average value. When $Y$   
changes from $R$ to $0$ in the ekpyrotic scenario,  $D(y, Y)$ for any   
$y>Y$ changes just few times. Indeed, consider the maximum and minimum   
values of $D(y,Y)$. At $y=0$ in ekpyrotic cosmology it takes the minimum   
value $D=C$. At $y=R$ it takes the maximum value $D(y, Y)= C + (\alpha   
-\beta)R + \beta Y$. Thus $ C \leq  D(y, Y)\leq C+\alpha R $. In the 
examples   
of \cite{KOST} $D_{min}=10^2$ and $D_{max}=3\cdot 10^2$ 
or $D_{min}=10^3$   
and $D_{max}=2\cdot 10^3$. Therefore, 
 the function $D$ changes 3 and 2 times in   
these two examples, respectively, so one can in the first approximation   
replace $D(y,Y)$ by its average value.   
   
On the other hand,  in the ekpyrotic cosmology  one had   
 $D \dot Y=  - \sqrt{ -2V(Y)}= - \sqrt {2v} e^{-\alpha m Y/2}$ \cite{KOST}.   
According to \cite{KOST,KKL}, when $Y$ changes from $R$ to $0$, the   
function $e^{\alpha m Y(t)/2}$  changes from some value greater   
than $e^{60}$ to $1$. This means  that during this time the scale factor   
$A(t) \sim {f(y)\over D^2\, \dot Y(t)}$ contracts at least $e^{30}$ times.

To compare this with the conclusions of \cite{KOST} note that they define   
the scale factor of the visible brane as   
\begin{equation}   
a_{-}(t) = A(t) \sqrt C\ ,   
 \label{visible}   
\end{equation}   
where $C$ is the constant in $D$. They conclude, using the 4d   
approximation, that $a_{-}(t)$ grows in time and the visible universe   
expands. But as we have seen, this contradicts strongly to the solution of   
the 5d dilaton equation in the bulk, which shows that $a_-(t)$   
exponentially contracts.   
   
However, the statement about expansion of the universe \cite{KOST}, as   
well as our statement about its exponential contraction, was based on   
averaging over the 5th dimension, assuming that the solutions of the 5d   
equations satisfying the metric ansatz of \cite{KOST} do actually exist.   
Now let us look at our exact result, Eq. (\ref{constraint}), more   
carefully. Since  $D(y,Y(t)) = (\alpha-\beta)y+C+\beta Y(t)$ is a function   
of both $y$ and $t$, one can easily check that this equation  is  formally   
inconsistent, i.e. it does not have any solutions at all! This simply   
means that the ansatz for the metric and the fields used in \cite{KOST}   
does not solve the 5d equations.

 One can come to a similar conclusion using   the 
gravitational equations  in the bulk. First, note that general covariance   
leads to a  relation between  5d Einstein equations,  dilaton and 4-form   
equations. One can verify that   the 4-form equation of motion in the   
bulk ${\delta S_{bulk}\over \delta A_{\alpha \beta \gamma \delta}}=0$ is   
satisfied by the time dependent ansatz of Ref. \cite{KOST}. Thus we get an   
identity:   
\begin{equation}   
\nabla^\gamma {\delta S_{bulk}\over \delta g_{\gamma \delta}}-   
\partial^\delta \phi {\delta S_{bulk}\over \delta \phi}=0  \ .   
 \label{generalcov}   
\end{equation}   
 For a solution of bulk equations of motion 
each term in this identity  should vanish.   
 However,  if the dilaton equation is not   
satisfied,  this identity  shows that a particular combination of the   
gravitational field equations cannot be satisfied by the ansatz   of 
\cite{KOST} (note that the derivatives $\partial^\delta \phi$ in the time   
and 5th direction do not vanish).   
   
To see  this more explicitly let us insert the ansatz of \cite{KOST}   
directly into the action, with $\dot B=\dot C=0$. We  shall change the   
variables so that   
\begin{equation}   
\tilde A^2(t,y, Y(t))=A^2(t)D^3(y, Y(t)) \ , \qquad \tilde N ^2(t,y,   
Y(t))= N^2(t)D^3(y, Y(t)) \ .   
 \label{variables}   
\end{equation}   
Then, up to total derivatives,   
\begin{equation}   
{\cal L}_{\rm bulk}(t,y)={M_3^2\over 2} \sqrt g\   
\left(R-\frac{1}{2}(\partial\phi)^2-{3\over 2}\frac{e^{2\phi}{\cal   
F}^2}{5!}\right) =  -{3 M_3^2 \tilde {A}^3 B\over \tilde N}   
 \left [\left({\dot {\tilde A} \over \tilde A}\right)^2   
 -{7 \over 4} \left({\dot D\over D}\right)^2 \right].   
\end{equation}   
Taking into account that  $\dot \phi= 3 \dot D/D$ we may  also rewrite the   
bulk action as   
\begin{equation} {\cal L}_{\rm   
bulk}(t,y)=  -{ M_3^2 \tilde {A}^3 B\over \tilde N}   
 \left [3 \left({\dot {\tilde A} \over \tilde A}\right)^2   
 -{7 \over 12} \dot \phi ^2 \right]. \label{Lbulk}   
\end{equation}   
   
This is a {\it local action} that should be varied to derive the {\it local   
equations}. If we perform the variation of this action over $\phi$, we   
will get the same dilaton equation as in (\ref{test1})  (note that $\tilde   
A$ and $\tilde N$ are independent of $\phi$), namely, $\partial_t({   
{\tilde A}^3 B\over \tilde N}\dot \phi)$=0 for $y>Y$.   
   
The reason why this equation  was not satisfied in \cite{KOST} is that   
they took the action of the form (\ref{Lbulk}), integrated it over $y$,   
added the brane action and neglected the term $J(Y) \beta^2 \dot   
Y^2(t)\sim \int dy \left( { M_3^2 \tilde {A}^3 B\over \tilde N}\dot   
\phi^2\right)$, as we have explained in the previous section (see eqs.   
(\ref{4daction}), (\ref{eff})). One may argue about whether this term is   
small or not as compared to the brane contribution in the integrated   
action. However, in 5 dimensions, i.e.  before the  integration over $y$,   
this term is given by ${ {\tilde A}^3 B \,  \dot \phi^2\over \tilde N}   
(t,y)$. For $y>Y$ there is no other contribution to the dilaton equation.   
Therefore the variation of this term must vanish as  Eq. (\ref{test})   
states.

\section{ Towards a consistent brane cosmology}\label{VY}   
   
As we have seen, different ways of taking averages in the situation where   
the 5d solutions do not exist can lead to dramatically different   
conclusions regarding  expansion versus contraction of the universe.   
Thus, if one really wants to investigate cosmological consequences of the   
ekpyrotic scenario, one should find exact solutions of the corresponding   
5d equations. This is especially important in the situation where the   
results of the averaging over the 5th dimension lead to the conclusions   
that are in an apparent contradiction with the Einstein equations in 4d,   
see Sect. \ref{4d}.

The fact that the ansatz for the metric and the fields used in \cite{KOST}   
does not solve the 5d equations is not very surprising.  Indeed,  it was   
shown in \cite{Binetruy:2000ut} that even in the simplest versions of   
brane cosmology describing  one or two branes one should use a  more   
general ansatz for the metric in order to satisfy the Israel junction   
conditions on the branes.  A generic metric which respects the planar   
symmetry of the problem has the form   
\begin{equation}\label{gener}   
ds^2= -n^2(t,y)dt^2+a^2(t,y)d{\vec x}^2+b^2(t,y)dy^2 \ ,   
\end{equation}   
Note that here the functions $a$, $b$, and $n$ depend both on $t$ and $y$, and there still is a residual freedom of transformation of the 
 coordinates $(t, y)$
Similarly, one may need to consider a more general ansatz for the fields  
as in \cite{Binetruy:2000ut}.

Before one begins looking for exact solutions using a more general metric   
ansatz, one should re-examine other assumptions of the theory. Indeed, in   
\cite{KOST} the potential $V(Y)$ was added by hand to the  bulk brane   
action, whereas the bulk  supergravity 
action remained unchanged. However, it is not   
obvious to us whether this is the proper way to introduce the inter-brane   
interactions in 5d.   
   
As an illustrative example, consider two charged plates of a capacitor in   
ordinary electrodynamics, positioned at $y = 0$ and $y = R$. If they have   
charges $q$ and $-q$, and the electric field between the plates is $E$,   
then the potential energy of the interaction between the plates can be   
represented as the ``brane potential'' $V(R) = -qE\,R$. However, it would   
be incorrect to think that this energy is localized on the plates. Rather   
it is concentrated in the electric field between the plates. It is possible to use the potential $V(R)$ to study
the motion of the branes. For example, if each brane has mass $M$, one can write $m\ddot R = -V'(R)$,
just as one does for the bulk brane acceleration in the ekpyrotic scenario. But if one studies gravitational backreaction of the electric field, it would be completely incorrect to replace the contribution of the electric field to
the energy-momentum tensor in the bulk by the delta-functional term proportional to $V(R)$.   
   
Similarly, if the potential $V(Y)$ appears due to the membrane instantons   
stretched between the branes, one should check whether the energy-momentum   
tensor in the bulk, as well as the dilaton and the 4-form field, changes   
due to these nonperturbative effects. Otherwise the appearance of the   
potential depending on the inter-brane separation would look as an example   
of the action at a distance.   
   
In a certain sense, the potential $V(Y)$ is analogous to the effective   
potential $V(r)$ of the radion field introduced by Goldberger and Wise   
\cite{GW}. It is a very useful concept if the only goal is to describe the   
forces acting on the bulk brane, ignoring the change of the metric   
produced by these forces. However, if one wants to study the corresponding   
changes in space-time geometry (and this was the main goal of   
\cite{KOST}), one should perform a full investigation of the inter-brane   
interactions in 5d \cite{DeWolfe:2000cp} and check whether one can add the   
fields responsible for the radion potential without additional fine-tuning   
and strong modification of the 4d geometry \cite{GKL}.

Thus  one has  a lot of things to do. First of all, one needs to find a   
theory with the potential $V(Y)$ which behaves as $-e^{-\alpha m Y}$ at   
large $Y$ (the functional form is important). This potential should be   
smaller in absolute value than $e^{-120}$ near the hidden brane, and   
should not have any positive contributions there with this accuracy. This   
fine-tuning is necessary to produce desirable density perturbations and   
avoid inflation. Also, this potential should not receive any contributions   
proportional to $e^{-\alpha m (R-Y)}$ due to the interaction with the   
hidden brane \cite{KKL}.  One must make sure that this potential vanishes at $y = 0$, to avoid the
cosmological constant problem. Then one must take into   
account that the visible brane and the hidden brane should be stabilized by   
some strong forces so that the effective potential of the corresponding   
moduli field could have mass  on the TeV scale or even  greater. One   
should also check that the strong forces leading to the brane   
stabilization do not interfere with the extremely weak interaction   
responsible for the potential $e^{-\alpha m Y}$. One cannot ignore the   
unresolved problem of brane stabilization (which was the position taken in   
\cite{KOST}) and speculate about the inter-brane potentials suppressed by   
a factor of $e^{-120}$.   
   
When/if the theory with the desirable $V(Y)$ is found, one should   
investigate its $5d$ nature and make sure that the effects producing   
$V(Y)$ do not induce large curvature on the branes. Then one should write   
down equations in 5d taking into account the energy-momentum tensor in the   
bulk together with the junction conditions, and solve them.   
   
And finally, one should find out what happens at the moment of the brane   
collision: whether the visible brane collapses, expands, stays at the same   
place or oscillates, etc. These issues have not been addressed in   
\cite{KOST}, and they cannot be fully analysed until the brane   
stabilization mechanism is understood.

Thus, if one wants to propose a consistent alternative to   
inflationary cosmology, one would need first to give a  consistent   
formulation of the alternative theory, and then find a correct solution of   
the corresponding equations. As we have seen, this is a rather nontrivial   
task.   
   
In this paper we studied a very limited part of this problem. We tried to   
check whether the basic assumptions of the ekpyrotic scenario (the ansatz   
for the metric and for the fields, and the modification of the bulk brane   
action proposed in \cite{KOST}) can lead to a consistent 5d description of   
an expanding visible brane. We have found that this is not the case.

\

The authors are grateful to T. Banks, K. Benakli, M. Dine, N. Kaloper, J.   
Maldacena, V. Mukhanov,  R. Myers, P. Nilles, A. Peet, K. Stelle, E. Witten for   
stimulating discussions  and to A. Van Proeyen for confirming the   
calculations of the 4-form. We thank NATO Linkage Grant 975389 for   
support. L.K. was supported by NSERC and by CIAR; L.K. and A.L. were supported by NATO Linkage Grant 975389; R.K. and A.L. and were   
supported by NSF grant PHY-9870115; A.L. was also supported by the   
Templeton Foundation  grant 938-COS273. A.T.  was supported by  the DOE   
grant DE-FG02-91ER40690, PPARC SPG grant 00613 and INTAS project 991590.

\section{Appendix: The 4-form story}

There are two points   
 about the 4-form dependence  in \cite{KOST} which must be changed   
to get  the  correct  set up for charged BPS 5d domain walls.   
 One has to  correct the   
coefficient in the action and  the coefficient in the solution for the   
4-form.     
These corrections in the form sector are important   
in order to test the time-dependent ansatz of \cite{KOST}.

\subsection{Action}   
   
The action     
in eq. (10) of \cite{KOST} is not the one to which they refer as given in   
\cite{lukas}, \cite{universe}. The one in \cite{universe}  does not have a   
4-form. We questioned the origin of their action in \cite{KKL} and they   
replied in \cite{Khoury:2001iy}:  {\it The 4-form formulation of the   
action is equivalent to the action presented in \cite{lukas,universe}.   
This is easily seen by eliminating the 4-form using its equation of   
motion}.

In \cite{universe} there were indeed the relevant  world-volume   
 terms in the   
action (only 2 branes at the fixed points are present there, so there   
$\beta=0$), but there were no Wess-Zumino terms with the 4-form field:   
\begin{equation}   
  S=-\frac{1}{2\kappa _5^2}\int _{{\cal M}_5} d^5x \sqrt{-g}\left[ R+\frac{1}{2V^2}\partial   
  _\alpha V\,\partial ^\alpha V+\frac{1}{3V^2}\alpha ^2\right]   
  -\sum_{i=1}^2 \frac{\sqrt{2}}{\kappa _5^2}\alpha _i\int _{{\cal M}_4^{(i)}}   
d^4  \xi_{(i)} \sqrt{-g}\, V^{-1} \,,   
 \label{StartAction}   
\end{equation}   
We take the constants in the bulk and boundary actions  as $\alpha   
=-\alpha ^{(1)}=\alpha ^{(2)}$. Then we make the redefinitions $   
  V\rightarrow e^\phi \,,\alpha \rightarrow \frac{3}{\sqrt{2}} \alpha   
  \,,\kappa _5^{-2}\rightarrow M_5^3 \,,  R\rightarrow -R\,   
 $   
so that our notation agree with \cite{KOST}. That leads to   
\begin{equation}
  S=\frac{M_5^3}{2}\int _{{\cal M}_5} d^5 x \sqrt{-g}\left[ R-\frac{1}{2}\partial
  _\alpha \phi \,\partial ^\alpha \phi -\frac{3}{2}e^{-2\phi }\alpha ^2\right]
  -3\sum_{i=1}^2 M_5^3\alpha _i \int _{{\cal M}_4^{(i)}} d^4  \xi_{(i)}
\sqrt{-g}\,e^{-\phi }\,.
 \label{StartActionredef}
\end{equation}  
The kinetic terms match the ones in  (10) in \cite{KOST}.      
Now we may perform the procedure suggested in \cite{BKV}. We promote the   
constant $\alpha$ to a function $G(x)$ and add a Lagrange multiplier  of   
the form $\varepsilon ^{\alpha \beta \gamma \delta \epsilon } (\partial   
_\epsilon   
 G)   
  {\cal A}_{\alpha \beta \gamma \delta }$ and assign the supersymmetry   
  transformation to the 4-form so that its variation will compensate the   
  variation of terms in the rest of the action with derivatives of $G$. If there are   
  no sources, from   
  the equation for the 4-form we learn that $G(x)$ is a constant, as   
  before in usual gauged supergravity in d=5 without a 4-form where there is a   
  constant gauge coupling. In   
the  presence of sources we will find that $G$ is piecewise   
  constant.   
  If there are  charged sources as in case of   
  \cite{universe} we may also add the WZ term.   
   
Thus we add a Lagrange multiplier term and a WZ term to the action of   
\cite{universe} given above in the form (\ref{StartActionredef}). Its   
normalization is arbitrary, we thus put a constant $c$ in front:   
\begin{equation}   
  S_A=\frac{c}{4!}\int _{{\cal M}_5} d^5 x   
\left[ -\varepsilon ^{\alpha \beta \gamma \delta \epsilon }   
  (\partial _\epsilon   
  G)   
  {\cal A}_{\alpha \beta \gamma \delta }+\sum_{i=1}^2 2\alpha _i\varepsilon ^{\mu \nu \rho \sigma }   
  {\cal A}_{\mu \nu \rho \sigma }\delta (x^5-x^5_i)\right] \,.   
 \label{SA}   
\end{equation}   
 The relative normalization between the two terms is   
arranged such that $\alpha $ jumps by 2$\alpha _i$ at brane $i$. If we   
differentiate over the 4-form we are back to (\ref{StartActionredef}). To   
have agreement with \cite{KOST} we have to choose $   
  c= {3\over 2} M_5^3.   
$ We may however do something else, namely, add and subtract a term   
quadratic in ${\cal F}$. We find   
\begin{equation}   
  S_G=   \int _{{\cal M}_5}  d^5x \left[ -\frac{3M_5^3}{4}\sqrt{-g} e^{-2\phi}   
  \left(G - {2 e^{2\phi}\over 3M_5^3 \sqrt{-g}}{c\over 5!}   
  \varepsilon ^{\alpha \beta \gamma \delta \epsilon }   
  {\cal F}_{\alpha \beta \gamma \delta \epsilon }\right)^2   
- \sqrt{-g}{ e^{2\phi}\over   
   5! 3M_5^3}c^2 {\cal F}^2   \right].   
\end{equation}   
Now we vary over the field $G$ and from  its equation we find that   
\begin{equation}   
  G = {2 e^{2\phi}\over 3M_5^3 \sqrt{-g}}{c\over 5!}   
  \varepsilon ^{\alpha \beta \gamma \delta \epsilon }   
  {\cal F}_{\alpha \beta \gamma \delta \epsilon } \ .   
\label{G-onshell}   
\end{equation}   
We thus find the action given in \cite{KOST} but where {\it the factor in   
front of the ${\cal F}^2$-term     
has to be changed by $3/2$.} The correct action (i.e.  the one which is   
equivalent to an action without 4-form in \cite{universe}) is   thus   
\begin{eqnarray}   
\nonumber & & S=\frac{M_5^3}{2}\int_{{\cal M}_5}   
d^5x\sqrt{-g}\left(R-\frac{1}{2}(\partial\phi)^2-{3\over   
2}\frac{e^{2\phi}{\cal   
F}^2}{5!}\right) \\   
& &-3\sum_{i=1}^3\alpha_iM_5^3\int_{{\cal M}_4^{(i)}}   
d^4\xi_{(i)}\Bigl(\sqrt{-h_{(i)}}e^{-\phi}   
 -\frac{\epsilon^{\mu\nu\kappa\lambda}}{4!}{\cal   
A}_{\gamma\delta\epsilon\zeta}   
\partial_{\mu}X^{\gamma}_{(i)}\partial_{\nu}X^{\delta}_{(i)}   
\partial_{\kappa}X^{\epsilon}_{(i)}\partial_{\lambda}X^{\zeta}_{(i)}\Bigr) \ .   
\label{eq:5daction1}   
\end{eqnarray}   
This action for  the two  branes at the fixed points (i.e. without  the   
third   
 brane) taken  in the   
static gauge with $X^\mu= \xi^\mu$ and $X^4 = Y=const$  does agree with   
the action  in \cite{universe}.   
   
\subsection{Bulk brane and 4-form}   
As was promised,  we will show here that the formulation of the 5d   
supersymmetric theory with a bulk brane between the orbifold planes is   
impossible without the use of the 4-form or, equivalently, without the   
gauge coupling field $G$ dual to the field strength $F=dA$ \cite{BKV}.

In the case of  two orbifold planes we may derive the action   
(\ref{StartAction}) by solving the equation of motion for the 4-form. This   
leads to   
\begin{equation}   
G'=   
 2\alpha \left(\delta(y) -\delta(y-R) \right)\,.   
 \label{G'}   
\end{equation}   
The solution for the gauge coupling field is $G(y)= \alpha \epsilon(y)$.   
It simply means that the gauge coupling is positive for positive $y$ and   
negative for negative $y$. Therefore,  $(G(y))^2= \alpha^2$ for all $y$   
and we find the bulk potential proportional to $e^{-2\phi} \alpha^2$ as it   
was given in the original form of the action (\ref{StartActionredef})   
where there was no 4-form, neither in the bulk, nor on the branes.   
   
Now  we may try to perform the same procedure of getting rid of the 4-form   
in case  when  the bulk brane is present. We find the gauge-coupling   
field dual to the 5-form field strength   
   
\begin{equation}   
G'=   
 2\alpha \left(\delta(y) -\delta(y-R) \right)- \beta \delta(y-Y) -\beta \delta(y+Y)\,.   
 \label{G'1}   
\end{equation}   
In  this case the solution for $G$ which is  dual to the 5-form takes   
values   
\begin{eqnarray}   
G^2 &=&\alpha^2    \hskip 2.8 cm   {\rm at} \qquad 0<|y|<Y \ ,\nonumber\\   
 G^2 &=&(\alpha-\beta)^2    \hskip 1.8 cm  {\rm at} \qquad   
Y<|y|<R \ .   
\end{eqnarray}   
If we would  try to plug this solution back into the action (into the term   
$e^{-2\phi}G^2$), we would find that the {\it bulk potential is a   
piecewise function, and,    therefore,  there is no local general   
covariant theory.}   
If, however,  we keep the 4-form, there exists a local general covariant   
5d bulk action. In the presence of sources,  this will lead to piecewise   
values of the 5-form for the solutions,  but the local general covariant   
action is available.

\subsection{Solution}   
   
Let us suppose   
 first that the action given in \cite{KOST}  were  correct and let us   try to   
check the solution for the 4-form. The  corresponding  equation is   
\begin{equation}   
  \partial_y  \left( \sqrt{-g}e^{2\phi }{\cal   
F}^{0123y}\right) +3\left[- \alpha\delta (y)+(\alpha -\beta )\delta (y-R)   
+{\beta\over 2}  \delta   
  (y-Y)+{\beta\over 2}  \delta   
  (y+Y) \right] =0\ .   
\label{Feq}   
\end{equation}   
For the ansatz in \cite{KOST}  we  have  $\sqrt{-g}=A^3BN[D(y)]^4$ while   
$\sqrt{-g_{(4)}}=A^3N [D(y)]^2$ and $e^{2\phi }=B [D(y)]^3$. We find  then   
that the solution is   
\begin{equation}   
{\cal F}_{0123y}= - {3\over 2} A^3  N B^{-1} D^{-2}(y)D'(y) \ ,   
 \label{wrong}   
\end{equation}   
where   
\begin{equation}   
 [D(y)]^{''}= 2 [\alpha\delta (y)-(\alpha -\beta )   
  \delta (y-R)] -\beta \delta (y-Y) -\beta \delta (y+Y) \ .   
\label{D1}   
\end{equation}   
{\it This differs from solution given in \cite{KOST} by a factor of $   
{3\over 2} A^3 N B^{-1}$} and by a sign. If we now start with   
the corrected  form of the action in eq. (\ref{eq:5daction})   
(with extra kinetic term factor $ {3\over 2}$),   
 we find  the corrected  expression for the 4-form   
\begin{equation}   
{\cal F}_{0123y}= - A^3  N B^{-1} D^{-2}(y) D'(y) \ .   
 \label{correct1}   
\end{equation}   
This solution differs from the one in \cite{KOST} by the sign and by the factor  $A^3  N B^{-1}$.

\end{document}